\documentclass{article}

\date{\today}

\newcommand{\Tr}{{\rm Tr}}
\newcommand{\insertplot}[5]{\begin{figure}
 \hfill\hbox to 0.05in{\vbox to #5in{\vfill
 \inputplot{#1}{#4}{#5}}\hfill}
 \hfill\vspace{-.1in}
 \caption{#2}\label{#3}
 \end{figure}}
 \newcommand{\inputplot}[3]{
 \special{ps: plotfile #1}
}
\newcounter{fig}   \newcommand{\lbfig}[1]{\refstepcounter{fig}
\label{#1} }

\usepackage{epsfig}
\usepackage{amsmath}
\usepackage{amsfonts}
\usepackage{graphicx}
\usepackage[german, english]{babel}
\usepackage{a4wide}
\usepackage{amsmath}
\usepackage{amssymb}
\usepackage{ifthen}
\usepackage{epsfig}

\begin{document}

\newcommand{\dd}{\mbox{d}}
\newcommand{\tr}{\mbox{tr}}
\newcommand{\la}{\lambda}
\newcommand{\ta}{\theta}
\newcommand{\f}{\phi}
\newcommand{\vf}{\varphi}
\newcommand{\ka}{\kappa}
\newcommand{\al}{\alpha}
\newcommand{\ga}{\gamma}
\newcommand{\de}{\delta}
\newcommand{\si}{\sigma}
\newcommand{\bomega}{\mbox{\boldmath $\omega$}}
\newcommand{\bsi}{\mbox{\boldmath $\sigma$}}
\newcommand{\bchi}{\mbox{\boldmath $\chi$}}
\newcommand{\bal}{\mbox{\boldmath $\alpha$}}
\newcommand{\bpsi}{\mbox{\boldmath $\psi$}}
\newcommand{\brho}{\mbox{\boldmath $\varrho$}}
\newcommand{\beps}{\mbox{\boldmath $\varepsilon$}}
\newcommand{\bxi}{\mbox{\boldmath $\xi$}}
\newcommand{\bbeta}{\mbox{\boldmath $\beta$}}
\newcommand{\ee}{\end{equation}}
\newcommand{\eea}{\end{eqnarray}}
\newcommand{\be}{\begin{equation}}
\newcommand{\bea}{\begin{eqnarray}}
\newcommand{\ii}{\mbox{i}}
\newcommand{\e}{\mbox{e}}
\newcommand{\pa}{\partial}
\newcommand{\Om}{\Omega}
\newcommand{\vep}{\varepsilon}
\newcommand{\bfph}{{\bf \phi}}
\newcommand{\lm}{\lambda}
\def\theequation{\arabic{equation}}
\renewcommand{\thefootnote}{\fnsymbol{footnote}}
\newcommand{\re}[1]{(\ref{#1})}
\newcommand{\R}{{\rm I \hspace{-0.52ex} R}}
\newcommand{\N}{{\sf N\hspace*{-1.0ex}\rule{0.15ex}%
{1.3ex}\hspace*{1.0ex}}}
\newcommand{\Q}{{\sf Q\hspace*{-1.1ex}\rule{0.15ex}%
{1.5ex}\hspace*{1.1ex}}}
\newcommand{\C}{{\sf C\hspace*{-0.9ex}\rule{0.15ex}%
{1.3ex}\hspace*{0.9ex}}}
\newcommand{\eins}{1\hspace{-0.56ex}{\rm I}}
\renewcommand{\thefootnote}{\arabic{footnote}}

\title{Skyrmion--Anti-Skyrmion Chains}

\author{
{\large Ya. Shnir}$^{\dagger \star}$
 and {\large D. H. Tchrakian}$^{\dagger\star}$ \\ \\
\\ $^{\dagger}${\small Department of Computer Science, National
University of Ireland Maynooth}
\\ {\small Maynooth, Ireland}
\\ $^{\star}${\small School of Theoretical Physics -- DIAS, 10
Burlington Road, Dublin 4, Ireland }}

\maketitle

\begin{abstract}
{\sf Static axially symmetric sphaleron-type solutions describing chains of interpolating
Skyrmion--anti-Skyrmions  have been constructed numerically. The configurations are
characterized by two integers $n$ and $m$, where $\pm n$ are the
winding numbers of the constituent Skyrmion and anti-Skyrmion and the second integer
$m$ defines type of the
solution, it has zero topological charge for even $m$ and for odd values of $m$ the Skyrmion--anti-Skyrmion
chain has topological charge $n$. For the vanishing mass term  we confirm the existence of such chain solutions
for winding number $|n| \ge 2$. The similarity with monopole--anti-monopole pairs is
highlighted.}
\end{abstract}


\section{Introduction}

The purpose of the present work is the construction of axially symmetric chain
lump--anti-lump solutions in the usual
Skyrme model~\cite{Skyrme:1961vq} on $\R^3$.
It was agrued over the last decade that the multi-monopole (MM) solutions of the Yang-Mills-Higgs model
and the multi-Skyrmion (MS) solutions of the Skyrme model have many features in common \cite{Houghton:1997kg}.
The question of constructing a Skyrme--anti-Skyrme pair (SAS)
follows naturally from the corresponding construction of a monopole--anti-monopole (MAP)
pair~\cite{Rueber,mapKK} and monopole--anti-monopole chains~\cite{KKS}, and like the former, it would
describe a sphaleron like configuration.
Likewise unsurprisingly, the resulting solutions are not
given in closed form but are constructed numerically.
The construction of an axially symmetric
Skyrme--anti-Skyrme pair was carried out recently by Krusch and Sutcliffe~\cite{Krusch:2004uf}, who pointed out to
its possible physical relevance as a model for the deuteron. In that case, with only a particle-antiparticle pair,
the axially symmetric solution would be expected to be of the minimal energy configuration, consistent with
its interpretation as a model for the deuteron.

This task was carried out using a gradient flow technique in \cite{Krusch:2004uf}. Given the considerable
numerical complexity of this problem relative to the corresponding monopole~\cite{mapKK,KKS} one, it is
in order to repeat it using a different numerical technique. This is done here, applying instead
a boundary value procedure. We extended the analysis presented in ref.~\cite{Krusch:2004uf} by inclusion of the
pion mass term in the Lagrangian of the model. Besides the Skyrme--anti-Skyrme pair solution, we constructed
new axially symmetric saddlepoint solutions which represent chains of $m$
single Skyrmions and anti-Skyrmions, each carrying
charge $n$ in alternating order. For an equal number of Skyrmions and anti-Skyrmions,
the chains reside in the topologically trivial
sector. When the number of Skyrmions exceeds the number of anti-Skyrmions by one, the chains reside in the sector
with topological charge $n$. The aim here is to reveal qualitative and quantitative similarities of chains of
Skyrmions and anti-Skyrmions, with those of chains of monopoles and antimonopoles. Unlike for the $m=2$ chain
in \cite{Krusch:2004uf}
however, we cannot expect that the axially symmetric chain configurations
with more numerous ($m \ge 2$) constituents
we construct are of the minimal energy.  Such sphaleron of the lowest energy may well exhibit discrete
symmetries.

The boundary value problem we consider applies to a two dimensional nonlinear partial differential equation
system and the numerical technique employed is exactly that applied in the corresponding monopole
problem~~\cite{mapKK,KKS}. The two dimensional system of equations is the result of the imposition of
{\it axial symmetry} to this problem. We recover of course the MS's carrying topological
charge $n$, which were found a long time ago for $n=2$
in \cite{Weigel:1986zc,Verbaarschot:1987au}, and for up to $n=5$ in \cite{Kopeliovich:1987bt}.

Before launching into the analysis, it should be pointed out that certain qualitative features expected
here are more similar to instanton solutions of the Yang--Mills system, rather than the corresponding
features of the monopole solutions of the Yang-Mills--Higgs system, inspite of the fact that the latter
are described on $\R^3$ like the Skyrmion, while the former are described on $\R^4$. There is a fundamental
qualitative difference between 'instanton' and 'monopole' type solutions, namely that the gauge connection
in the former is asymptotically {\it pure gauge} while in the latter case it is {\it one half pure gauge}.
It turns out that the asymptotic properties of the solitons of sigma models are 'instanton' like,
irrespective of dimensions, in that the {\it composite connection} is asymptotically {\it pure gauge}.
Thus we expect from the outset, that the Skyrme--anti-Skyrme lumps should be more akin to the
instanton--anti-instanton~\cite{Radu:2006gg}, rather than the monopole--anti-monopole. The property we
have in mind is that there exists a monopole--anti-monopole consisting of a charge $n$ and a charge $-n$ pair
with $n=1$, and, with winding number $n>1$~\cite{Paturyan:2003hz}. It turns out that the corresponding
instanton--anti-instanton exists only for values $n\ge 2$. This was found in \cite{Radu:2006gg} only as a
result of a numerical construction, which is in agreement with the same nonexistence result for $n=1$ in
the analytic proof for the existence of such nonselfdual lumps given in \cite{Sadun:1992vj,Bor:1992ai}. Thus
we would not expect here, to find a Skyrme--anti-Skyrme lump consisting of a charge $1$ and charge $-1$
pair. We do however find such a pair when the pion-mass potential in introduced.
We will return to elaborate on this feature of the solutions that we find, in our conclusions.
In section $2$ we impose axial symmetry and calculate the resulting reduced two dimensional energy density
functional whose second order equations will be integrated, as well as the topological (baryonic) charges of
these configurations. Our numerical results are presented in section $3$, and our conclusions in
section $4$.

\section{The model and imposition of symmetry}
In terms of the order parameter multiplet $\phi^a=(\phi^{\al},\f^3,\f^4)$, $\al=1,2$, of the
nonlinear $O(4)$ sigma model field subject to $|\f^a|^2=1$, the rescaled static Hamiltonian of the
Skyrme~\cite{Skyrme:1961vq} model is expressed as
\be
\label{model}
{\cal H}_{\rm stat}=|\pa_i\f^a|^2+\frac{1}{4}|\pa_{[i}\f^a\pa_{j]}\f^b|^2 + \mu^2 (1-\phi^3)
\ee
$i=x,y,z$, with the notation $[ij]$ implying antisymmetrisation.
Here $\mu^2 (1-\phi^3)$ is the pion mass term.

The energy density \re{model} is bounded from below by the topological charge density
\bea
\varrho_0&=&\frac{1}{24\pi^2}\,\vep_{ijk}\vep^{abcd}\pa_i\phi^a\pa_j\phi^b\pa_k\phi^c\phi^d
\nonumber\\
&=&\frac{1}{24\pi^2}\,\vep_{ijk}\vep^{\al\beta\gamma}(\pa_i\phi^{\al}\pa_j\phi^{\beta}
\pa_k\phi^{\gamma}\phi^4-
3\pa_i\phi^{\al}\pa_j\phi^{\beta}\pa_k\phi^4\phi^{\gamma})\,,
\label{baryon}
\eea
whose integral is the integer winding number $n$, namely the baryon number. It is well known that this
lower bound cannot be saturated and hence we are concerned only with the second order Euler-Lagrange
equations.

Next, we state the axially symmetric Ansatz
parametrised by two functions $f=f(\rho,z)$ and $g=g(\rho,z)$, $\rho^2=x^2+y^2$,
and in terms of the two component unit vector $n^{\al}$
imposing axial symmetry on the $O(4)$ sigma model
field $\phi^a=(\phi^{\al},\phi^3,\phi^4)$
\be
\label{n}
n^{\al}=
\left(\begin{array}{c}
\cos n\vf \\
\sin n\vf
\end{array}\right)
\ee
with the integer $n$ counting the winding of the azimutal angle $\vf$, being the baryon number.

This Ansatz is expressed compactly as
\bea\label{ansatz}
\f^{\al}&=&\sin f\sin g\,n^{\al}\equiv a(\rho,z)\,n^{\al}\nonumber\\
\f^3&=&\sin f\cos g\equiv b(\rho,z)\label{axial}\\
\f^4&=&\cos f\equiv c(\rho,z)\nonumber
\eea
with the functions $f=f(\rho,z)$ and $g=g(\rho,z)$ dependent on the radial coordinate
$\rho=\sqrt{|x_{\al}|^2}$ of the $\R^2$ subspace, with the index $i=1,2$, and $z=x_3$.
$\hat x^{\al}=x^{\al}/{\rho}$ is the unit radius vector in the $\R^2$ subspace and $n^{\al}$ is the unit vector with
vortex number $n$. The Ansatz (\ref{ansatz}) has the same general structure as the parametrization
used in \cite{Krusch:2004uf}.  Furthermore,
one readily verifies that the parametrisation (\ref{ansatz}) is consistent, i.e., the complete
set of the field equations which follows from the variation of the original action of the Skyrme model, is
compatible with two equations which follow from variation of the reduced action on the Ansatz
(\ref{ansatz}).
In discussing the asymptotics we employ polar coordinates replacing $r=\sqrt{\rho^2+z^2}$,
$\ta=\arctan\frac{z}{\rho}$. It is also convenient to use the
trigonometric parametrization of the Skyrme model in terms of the functions
$f,g$ to represent the energy functional of the model and the topological charge (baryon number) density
\cite{Sawado:2004yq}, although it is not appropriate from the point of view of numerical calculations because of
the numerical errors which originate from the disagreement between the boundary conditions on the angular
type function $g(r,\theta)$
on the $\rho$-axis and the boundary points $r = 0, \infty$, respectively\footnote{For the MS
solution it results in the appearance of step function like dependence at these points: $g(0, \theta) =
g(\infty, \theta) = \pi~\Theta(\theta-\pi/2)$, $\Theta$ being the step function.}.

Indeed, the reduced two dimensional energy density functional resulting from the imposition of axial symmetry
stated in the Ansatz \re{ansatz}, is given by \cite{Sawado:2004yq},
\bea \label{engred-trig}
E &=& \frac{1}{r^2} \biggl((r \partial_r f)^2 +  (\partial_\theta f)^2 +
\left[ (r \partial_r g)^2 + (\partial_\theta g)^2 + \frac{n^2 \sin^2 g}{\sin^2 \theta} \right]\sin^2 f
\biggr) \nonumber \\
&+& \frac{\sin^2 f}{r^4} \biggl( (r \partial_r f \partial_\theta g -
r \partial_r g \partial_\theta f )^2 + \frac{n^2 \sin^2 g}{\sin^2\theta}
\biggr[ (r \partial_r f)^2 + (r \sin f \partial_r g)^2 \nonumber\\
&+&   (\partial_\theta f)^2 +
  (\sin f \partial_\theta g)^2 \biggr]  \biggr) + \mu^2(1-cos f)
\eea
or, equivalently
\bea \label{engred}
E &=& \biggl[1 +  \left( \frac{na}{\rho}\right)^2 \biggr]\biggl[
(a_\rho^2 + b_\rho^2 + c_\rho^2) + (a_z^2 + b_z^2 + c_z^2)\biggr] \nonumber \\
&+&  \biggl( \frac{na}{\rho}\biggr)^2 + \biggl( (a_{[\rho}\,b_{z]})^2
+ (b_{[\rho}\,c_{z]})^2 + (c_{[\rho}\,a_{z]})^2 \biggr) + \mu^2(1-c) \eea
where $a_{[\rho}\,b_{z]}\equiv \partial_\rho a \partial_zb - \partial_\rho b \partial_z a = a_\rho b_z - b_\rho a_z$.
The latter truncated functional represents some modication of the $O(3)$ sigma model on the half-plane.

We shall be choosing our boundary conditions such that the resulting multi-Skyrmion, or, the Skyrmion--anti-Skyrmion
chain solution has the appropriate baryon charge. In the case of the multi-Skyrmions this is the (topological)
winding number $n$ appearing in \re{n} and \re{ansatz}. In the case of Skyrmion-antiSkyrmion chains with an odd number
of lumps the baryon number is again $n$, and it vanishes for Skyrmion--anti-Skyrmion chains with an even number
of lumps. The topological charge is labeled with two distinct integers, the winding number $n$ and a second integer
$m$ which specifies the asymptotic value of the function $g(r,\ta)$ in \re{ansatz} when $r\to\infty$, namely
\be
\label{m}
\lim_{r\to\infty}g(r,\ta)=m\,\ta\,.
\ee
The formula for the baryon number that we use is
\be
\label{volinttt}
B = \int\varrho_0d^3x=-\frac12\,n\
\left[\cos m\theta\right]_{\theta=0}^{\theta=\pi}=\frac12\,n\,\left[1-(-1)^m\right]\,.
\ee
the configurations being classified according to the values of two integer numbers, $n$ and $m$.
The consideration above indicates that
the case $m=1$ corresponds to the (multi)skyrmions of topological charge $n$, while
$m=2$ yields a configuration with zero net topological charge consisting of two lumps; a
sphaleron like axially symmetric static solution  of the Skyrme model,
consisting of a charge $n$ Skyrmion and charge $-n$ anti-Skyrmion. More general,
for odd values of $m$
the winding number $n$ coincides with the topological charge of the Skyrme field $B=n$ whereas even values of $m$
correspond to the deformations of the topologically trivial sector. In the following we shall see that the
value of the integer $m$ defines the number of the constituents of the configuration which can be identified with
individual charge $n$ Skyrmions and charge $-n$ antiSkyrmions placed along the axis of symmetry
in alternating order.

\section{Numerical results}
The Euler-Lagrange equations arising from the variations of \re{engred}
have been integrated by imposing the  boundary
conditions, which respect finite mass-energy and finite energy density
conditions as well as regularity and symmetry requirements. Also the sigma-model constraint
is imposed.

The numerical calculations are performed employing the package FIDISOL/CADSOL,
based on the Newton-Raphson iterative procedure \cite{FIDISOL}.
We solve the system of three coupled nonlinear partial differential equations
numerically, on a non-equidistant grid in
$x$ and $\theta$,
employing the compact radial coordinate
$x= r/(1+r) \in [0:1]$. Typical grids used have sizes $75 \times 60$.

Note the parametrization in terms of the angular type functions $f(r,\theta),~{\rm and}~ g(r,\theta)$
is plagued by the disagreement between the boundary conditions we have to impose on the function
$g(r,\theta)$ on the $\rho$-axis and the boundary points $r = 0, \infty$, respectively. Therefore
it will be correct to impose boundary conditions on the fields $a,b,c$ as \cite{Krusch:2004uf}
\begin{eqnarray}
\label{infty}
a|_{r=\infty}=0\,,~~b|_{r=\infty}=0\,,~~c|_{r=\infty}=1
\end{eqnarray}
at infinity, and for the multi-Skyrmions and Skyrmion--anti-Skyrmion chains with odd number of constituents
we require
\be
\label{orig1}
a|_{r=0}=0\,,~~ b|_{r=0}=0\,~~ c|_{r=0}=-1,
\ee
at the origin
\footnote{Note that the boundary conditions \re{orig1}
correspond to the negative values of the baryon number.}.
For the Skyrmion--anti-Skyrmion
pair and the chains with even number of constituents
the Neumann boundary conditions must be imposed there on the fields $b,c$, i.e.,
\be
\label{orig2}
a|_{r=0}=0\,,~~ \partial_r b|_{r=0}=0\,~~ \partial_r c|_{r=0}=0,
\ee

The boundary conditions along the $z$-axis for odd and even values of the integer
number $m$ appearing in \re{volinttt} are
\begin{eqnarray}
\label{tapi2MS}
 a|_{\theta=0}=0, &~~& \partial_\theta b|_{\theta=0}=0,~~\partial_\theta c|_{\theta=0}=0;\nonumber \\
 a|_{\theta=\pi}=0, &~~& \partial_\theta b|_{\theta=\pi}=0,~~\partial_\theta c|_{\theta=\pi}=0
\end{eqnarray}

To produce a configuration with correct topology and boundary condition which can be used as an input
into the system of Euler-Lagrange equations,
we implemented following algorithm. The trigonometric parametrisation of the triplet $a,b,c$ given by
the Ansatz (\ref{ansatz}) is used with linear dependence of the profile function $f(x,\theta) = \pi(1-x), ~x\in [0,1]$
on the compact radial variable $x$ and linear dependence of the second angular
function $g(x,\theta) = m \theta, ~\theta \in [0,\pi]$  on the polar angle $\theta$. Note we do not impose additional
symmetry restrictions on the triplet of fields  $a,b,c$ in the input configuration, however the resulting
numerical solutions reveal such a discrete symmetry with respect to the reflection $z \to -z$. In addition to the
solution of the boundary problem performed with the package FIDISOL/CADSOL,
the gradient flow equations are solved to check the numerical results although in the latter case convergence is
somewhat slower.

As a first step we reproduced the well known results for charge $n$ multi-Skyrmions
\cite{Weigel:1986zc,Kopeliovich:1987bt,Verbaarschot:1987au} and for the charge 2 Skyrmion and charge -2 anti-Skyrmion
pair \cite{Krusch:2004uf}.
Evidently, this algorithm can be implemented to construct Skyrmion--anti-Skyrmion chains, the
configurations similar to the monopole-antimonopole chains constructed in
\cite{Rueber,mapKK,KKS,Paturyan:2003hz}.
For example, the value of the integer number $m=3$ corresponds to the charge $-n$
anti-Skyrmion located at the origin and two charge $n$ Skyrmions located on the $z$-axis
symmetrically with respect to the $xy$-plane, the system with $m=4$, which
resides in the topologically trivial sector, corresponds to the
chain of 2 SAS pairs alternating on the symmetry axis, etc.

Indeed, for the values of the winding number $m \ge 3$ in the input configuration we
produce the Skyrmion-anti-Skyrmion chain solutions with winding numbers $n\ge 2$.
The relative error is estimated to be lower than $10^{-3}$. Another check of the
correctness of our results was performed by verifying that the virial relation~\cite{taubes},
namely the identity that ensues from the Derrick scaling requirenment
\footnote{In this case, the identity is $T=G-3V$, consisting of the positive definite
integrals $T = \int d^3x |\partial_i \f^a|^2 $,
$G=\frac14 \int d^3x  |\pa_{[i}\f^a\pa_{j]}\f^b|^2$
and $V=  \mu^2 \int d^3x (1-\phi^3)$}
, is satisfied.  This was done both for multi-Skyrmions (MS's) of different topological charges
and for Skyrmion-antiSkyrmion chain configurations.

Although the initial configuration
satisfies the boundary condition  \re{m}, we do not impose it as a boundary condition in our
numerical calculations, nevertheless the solutions  asymptotically tend to satisfy \re{m}.

The numerical algorithm we impemented to solve the boundary problem allows us to evaluate the dipole moment
of the solutions. Indeed, the leading term in the asymptotic expansion of the axially symmetric
multi-Skyrmion solution is a dipole \cite{Manton:1994ci}, i.e.,
$$
\f^3 \sim  \frac{d \cos \theta}{r^2} + O(r^{-3})\, ,
$$ so in terms of the compact radial coordinate $x$ we can extract the value of the dipole moment from the
first and second derivatives of the field $\f^3$ at the boundary $x=1$:
$$
d = \partial_x \f^3(\theta,x)\biggl. \biggr|_{x=1, \theta=0} +
\frac{1}{2}\partial_{xx}^2 \f^3(\theta,x) \biggl.\biggr|_{x=1, \theta=0}
$$
\begin{figure}
\lbfig{f-1}
\begin{center}
\includegraphics[height=.30\textheight, angle =-0]{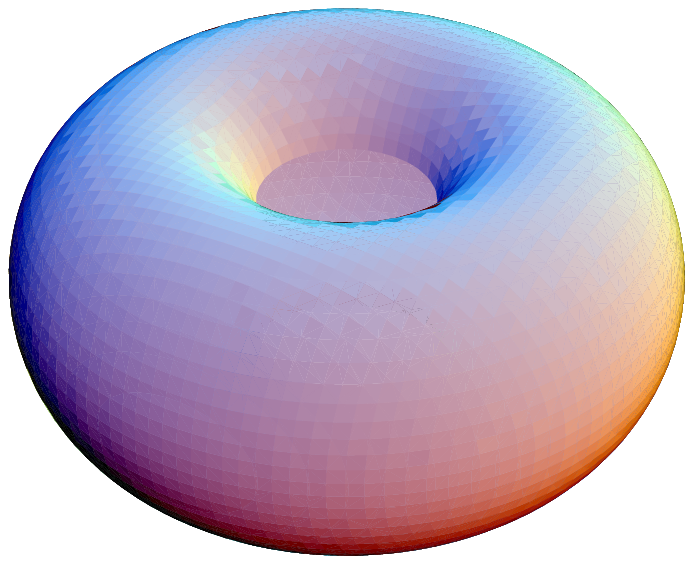}
\includegraphics[height=.30\textheight, angle =-0]{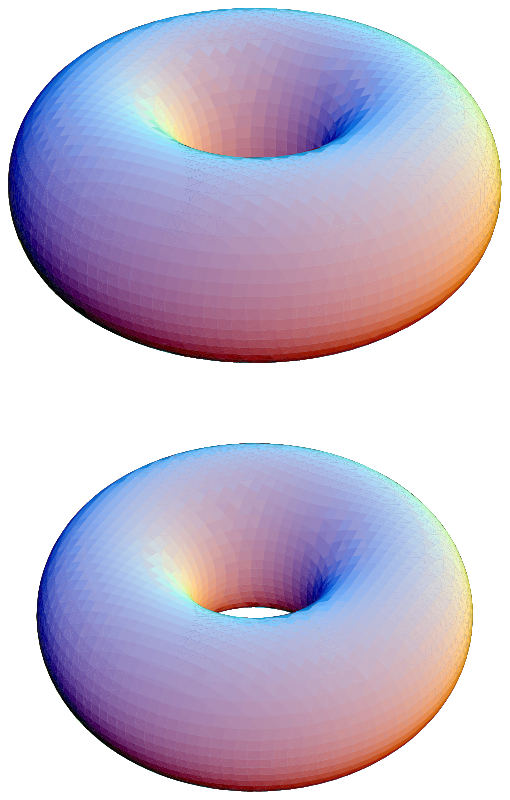}
\includegraphics[height=.30\textheight, angle =-0 ]{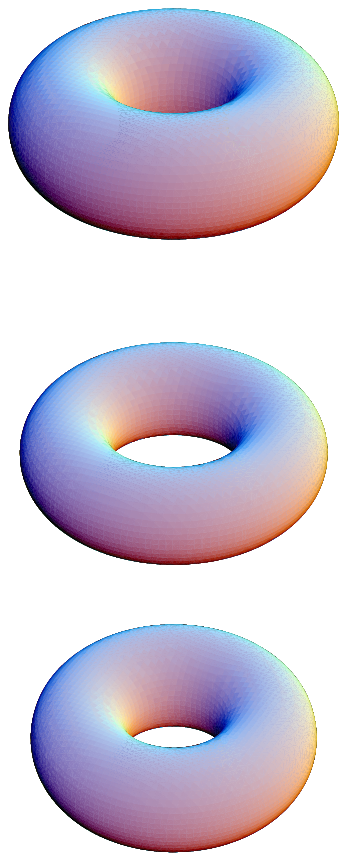}
\includegraphics[height=.36\textheight, angle =-0]{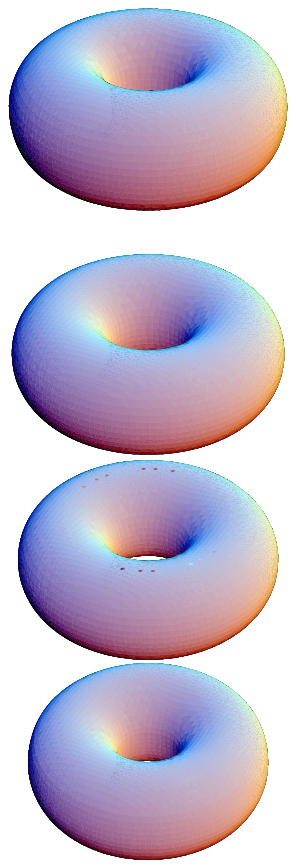}
\includegraphics[height=.36\textheight, angle =-0 ]{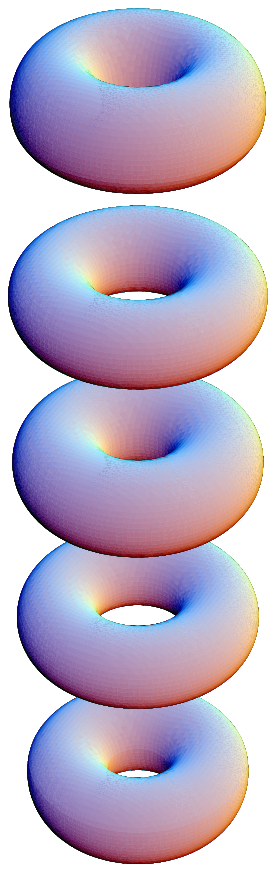}
\includegraphics[height=.36\textheight, angle =-0 ]{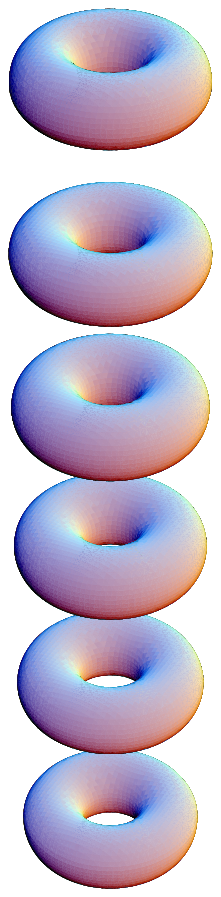}
\end{center}
\caption{The 3d energy isosurfaces of the charge 2 axially symmetric Skyrmion and different
Skyrmion--anti-Skyrmion chains for $n=2$, $m=1, \dots 6$ are shown for $\mu = 0$.
(in different scales, cf. Table 1).}

\end{figure}
\noindent
Fig. \ref{f-1} displays the energy density surfaces for the charge 2 axially symmetric Skyrmion
($n=2, m=1$),  charge 2 Skyrmion and charge -2 anti-Skyrmion
pair ($n=2, m=2$) and for the Skyrmion--anti-Skyrmion chains where Skyrmions and anti-Skyrmions
alternate along the symmetry axis ($n=2, m=3,4,5,6$). Positions of the constituents can be identified as points in space
where the field $\f^4 \equiv c(\rho_0,z_0)$ is equal to $-1$ \cite{Krusch:2004uf}. Indeed, Fig \ref{f-2}
demonstrates that these points
almost coincide with the maxima of the energy density distribution.
\begin{figure}
\lbfig{f-2}
\begin{center}
\includegraphics[height=.30\textheight, angle =-90]{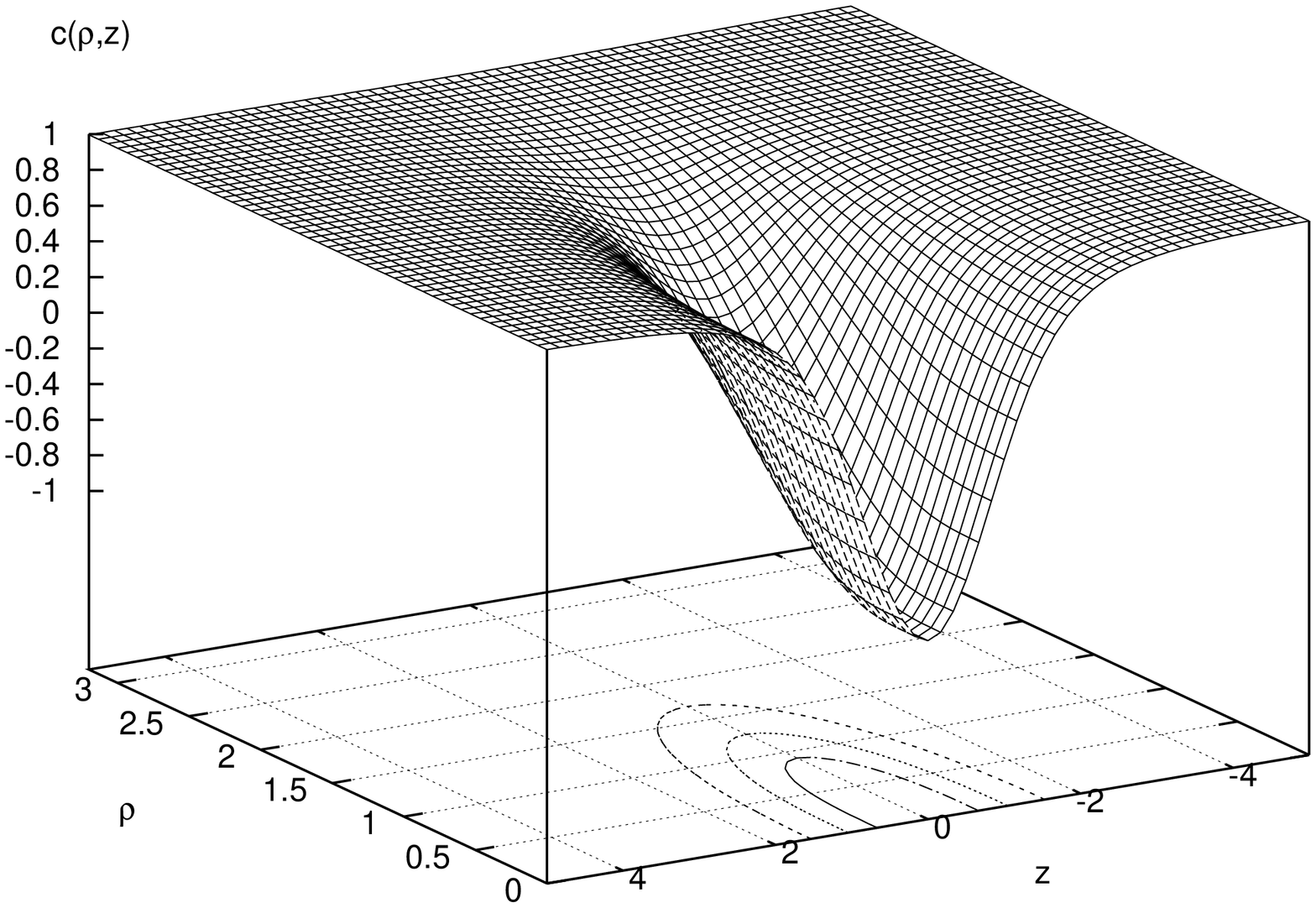}
\includegraphics[height=.30\textheight, angle =-90]{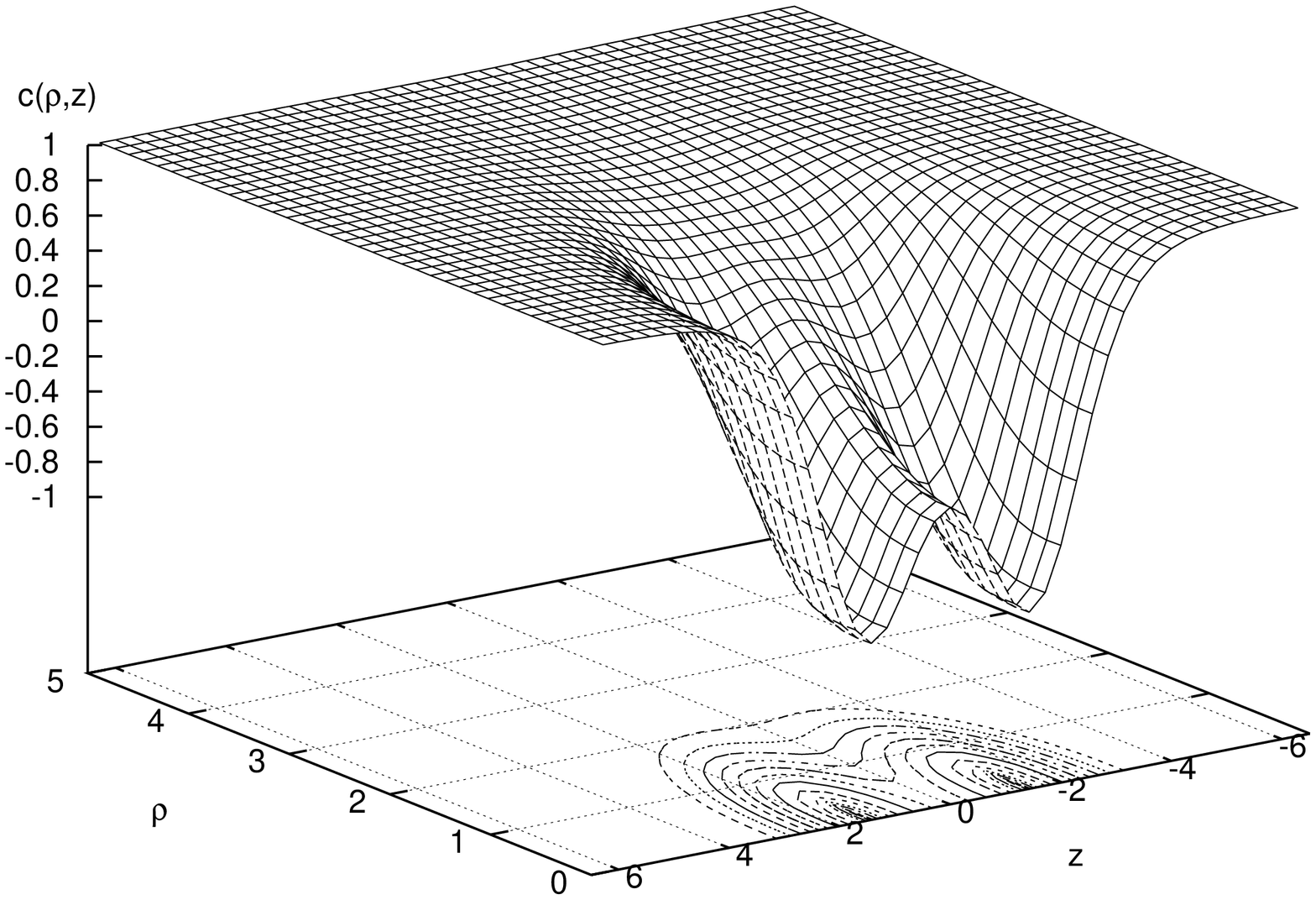}
\includegraphics[height=.30\textheight, angle =-90 ]{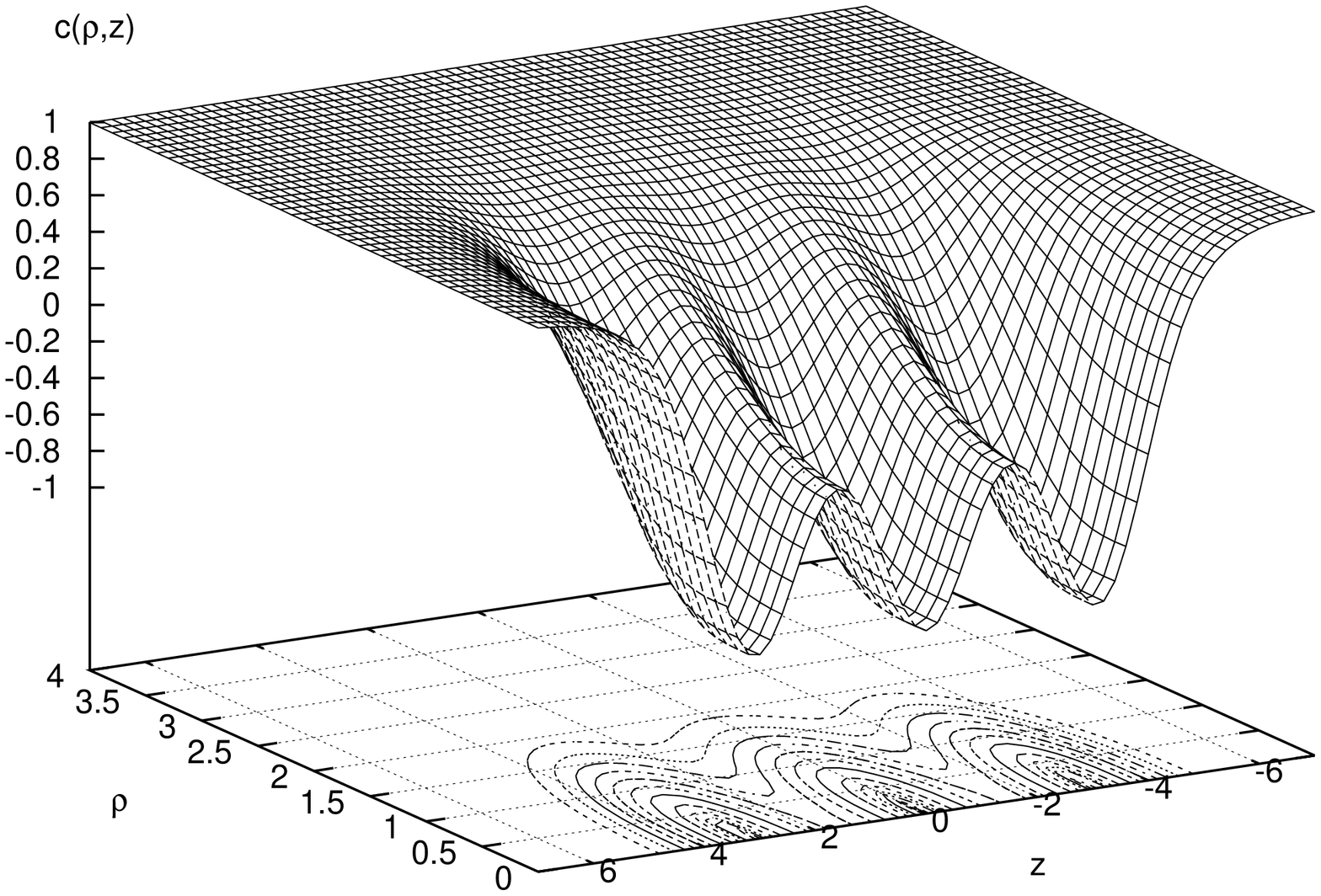}
\includegraphics[height=.34\textheight, angle =-90]{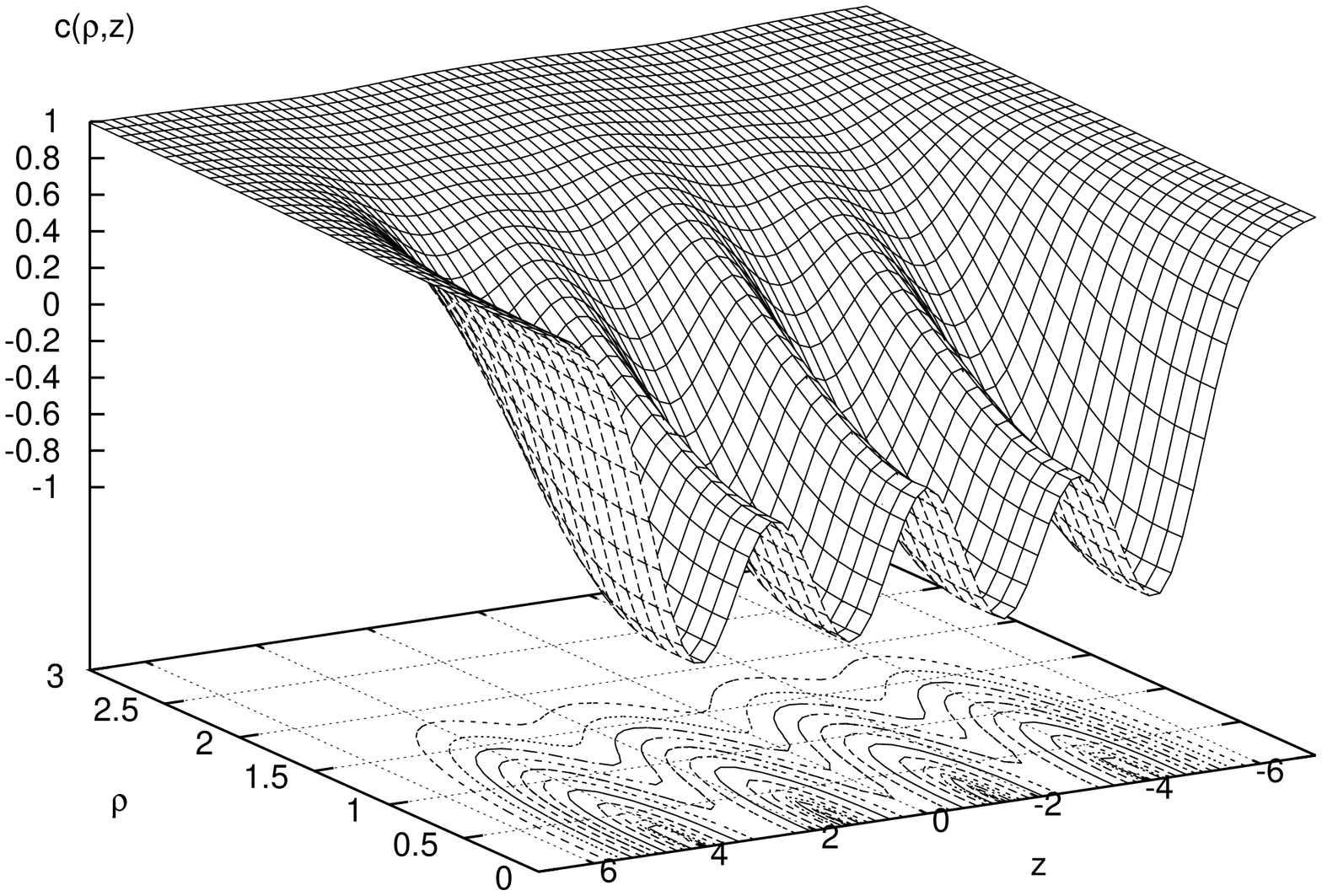}
\includegraphics[height=.32\textheight, angle =-90 ]{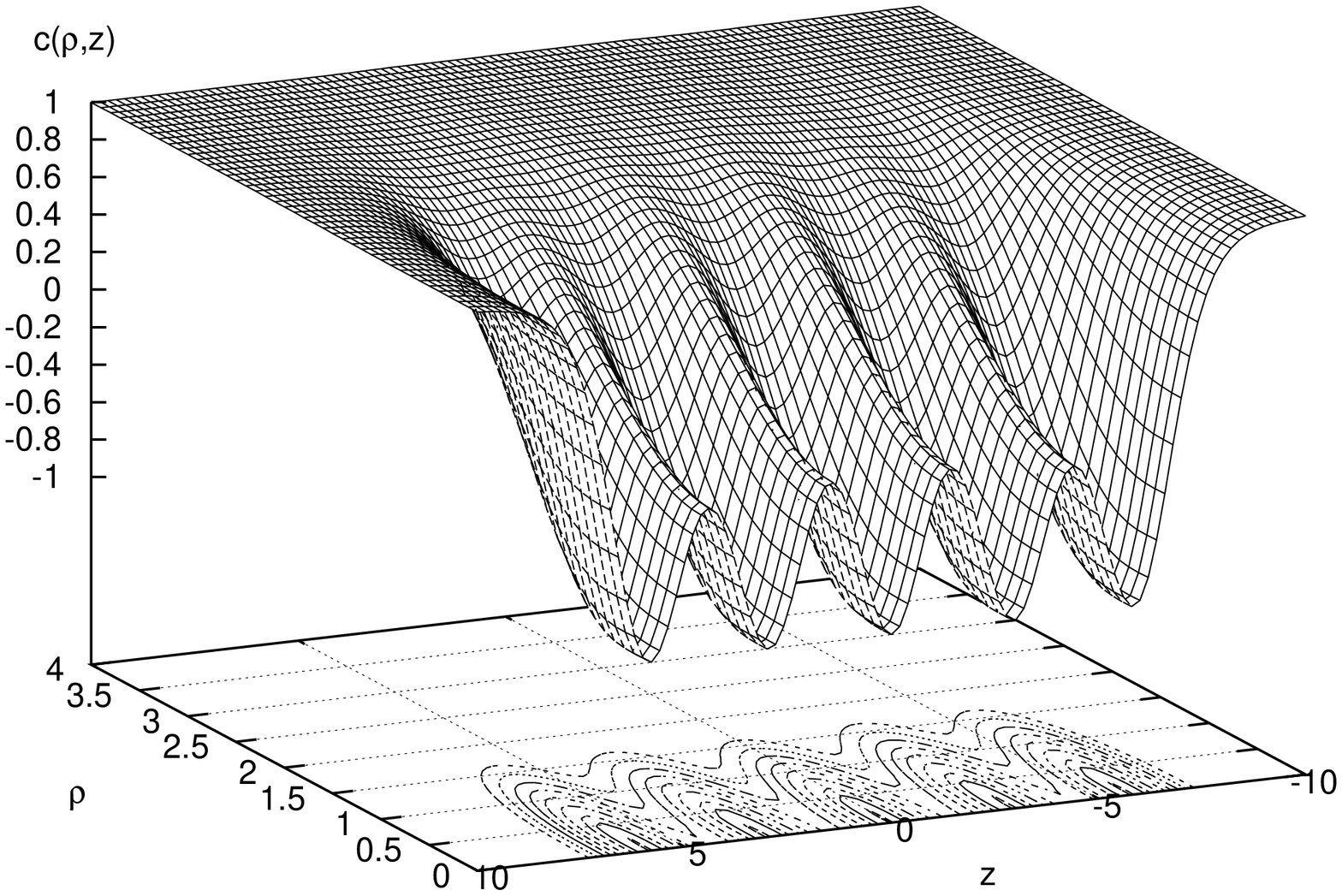}
\includegraphics[height=.32\textheight, angle =-90 ]{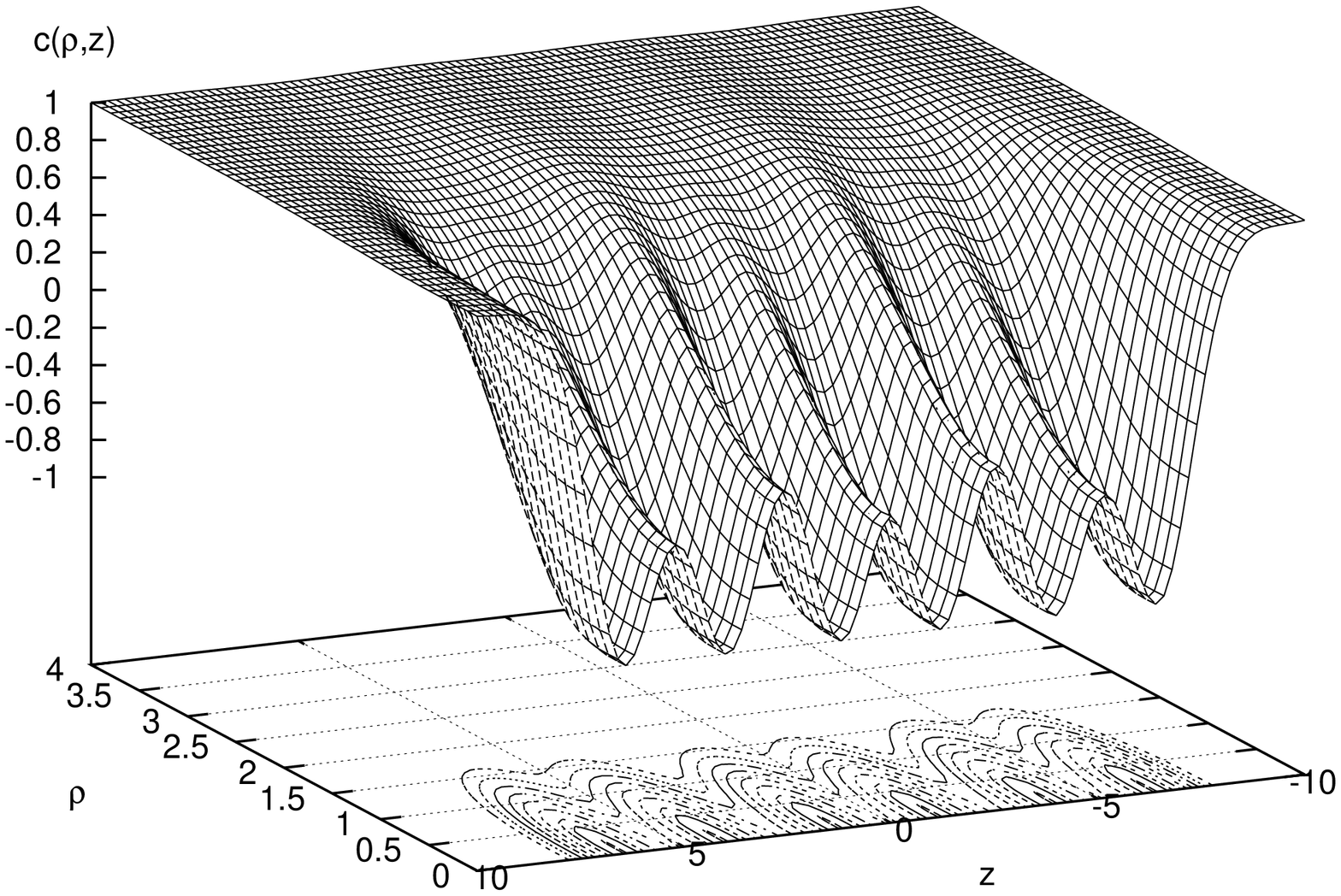}
\end{center}
\caption{
The field function $c(\rho,z)$ of the $m=1$ charge 2 Skyrmion and Skyrmion-anti-Skyrmion chains
for $n=2$, $m=1, \dots 6$ are shown for $\mu = 0 $ as functions of the coordinates $\rho,z$.
}
\end{figure}
The higher energy Skyrmion--anti-Skyrmion chains are formed from $m$ constituents of charge $\pm n$.

As an example, in Table 1 we present the energy $E$ of the Skyrmion--anti-Skyrmion chains with winding number $n=2$,
the interaction energy, which is defined as $  \Delta E =  m E^{(0)} - E \ ,$ where $E^{(0)} = 2.362$ is the energy
of single charge $n=2$ Skyrmion, the loci $z_i$ of
the field $c(0,z_i)$ and the numerical values of the dipole moment $d$ for the solutions with
$\theta$ winding number $1\leq m \leq 6$.

\begin{center}
\begin{tabular}{|c|c|c|c|c|}
 \hline
 m  &  $E$  & $\Delta E$ & $z_i$ &$d$   \\
 \hline
1   & $2.36$&   & $0.0$ & $4.31$      \\
2   & $4.64$& $0.08$  & $\pm 1.44$ & $0$   \\
3   & $6.94$ & $0.14$ & $0.0 $ $\pm 3.17$ & $4.31$  \\
4   & $9.26$ & $0.19$ & $\pm 1.54 \ $ $\pm 4.45$& $0$   \\
5   & $11.57$& $0.24$  & $0.0 \ $ $\pm 3.12 \ $ $\pm 6.05 $ &$4.28$   \\
6   & $13.88$& $0.29$  &$\pm 1.62 \ $ $\pm 4.57 \ $ $\pm 7.42$ & $0$ \\
 \hline
\end{tabular}\vspace{7.mm}\\
{\bf Table 1}
The energy of the Skyrmion-anti-Skyrmion chains, the
positions of the constituents $z_i$ and the values of the dipole moments
are given for the $n=2$ $m$-chains with $m=1,\dots ,6$ for $\mu = 0.$
\vspace{7.mm}\\
\end{center}

A qualitative similarity between these Skyrmion--antiSkyrmion chains and the monopole--anti-monopole chain solutions
~\cite{KKS} is that in both cases there is a picture of an effective interaction between the constitutents which
allows the sphaleron-type solution to exist, although the nature of the interaction is different. In the former case
it is an effective electromagnetic interaction between the constituents \cite{KKS,Shnir:2005te} while in the latter case
there is a dipole-dipole interaction between the Skyrmions \cite{Manton:1994ci}.
We observe that the distances between positions of the constituents do not vary much within a chain. One can
try to model the Skyrmion--antiSkyrmion chains  in the
framework of the effective dipole-dipole interaction of the Skyrmions as was conjectured in \cite{Krusch:2004uf}.

Indeed the numerical results
suggest the dipole moment of the Skyrmion-antiSkyrmion pair associated with the asymptotic behaviour
of the field $b$, is vanishing. We observe similar results for
chains with even number of constituents. For odd values of $m$ there is only one component of the
dipole moment of the configuration, directed along the symmetry axis. For these  chains it almost coincides
with the dipole moment of the single charge $n$ Skyrmion. Considering multiSkyrmions of charge $n$, we can verify
the conjecture of \cite{Krusch:2004uf} concerning the additivity of the dipole moments. Indeed for the axially symmetric
Skyrmions with topological charge  $n=1, \dots ,4$ the dipole moment behaves approximately as $d=n d_1$ where
$d_1 = 2.16$ is the dipole moment of the spherically symmetric charge one Skyrmion, although the results of the
numerical calculations (cf. Table 2)
indicate it is slightly higher than this additive estimate for $n\ge 4$. On the other hand it is
known that the axially symmetric configurations of charge $n \ge 3$ are not the minimal energy states of the system
\cite{Houghton:1997kg}.
\begin{center}
\begin{tabular}{|c|c|c|c|c|c|c|}
 \hline
$n$  &  $1$  & $2$ & $3$ &4 &5 &6   \\
 \hline
$d$  & $2.161$&  $4.31$ & $6.69$ & $9.47$ & $12.64$ & $16.23$      \\
$M$  & $1.24$&  $2.36$ & $3.57$ & $4.83$ & $6.14$ & $7.47$      \\
 \hline
\end{tabular}\vspace{7.mm}\\
{\bf Table 2}
The values of the dipole moments of the $m=1$ multiSkyrmions with topological charge  $n=1, \dots ,6$
and their masses are given for $\mu = 0.$
\vspace{7.mm}\\
\end{center}
Note that both types of the chain solutions with even and odd values of $m$
possess the symmetry
$$
(a(\rho,z), b(\rho,z), c(\rho,z)) \to (-a(\rho,z), -b(\rho,z), c(\rho,z))
$$
which corresponds to the inversion of the asymptotic pion dipole fields.

For $n=2,3,4$ these $m$-Skyrmion chains possess $m$ points on the $z$-axis where the field
$\f^4(0,z_0) = c(0,z_0)=-1$ and a soliton is placed. Due to reflection symmetry, each such a point $-z_0$ on the
negative $z$-axis corresponds to a point $z_0$ on the positive $z$-axis. For even values of $m$ the triplet of the
fields $a,b,c$ has reflection symmetry with respect to the $xy$ plane
$$
(a(\rho,z), b(\rho,z), c(\rho,z)) \to (-a(\rho,-z), b(\rho,-z), c(\rho,-z))
$$
while for odd values of of $m$ the reflection symmetry is
$$
(a(\rho,z), b(\rho,z), c(\rho,z)) \to (a(\rho,-z), -b(\rho,-z), c(\rho,-z))
$$
Note that these symmetries are not imposed by the boundary conditions on the fields but arise
when an initial configuration relaxes into a solution.
For even $m$ the field $c(0,0)$ is far from the value $-1$ at the origin, although
its value there decreases with increasing of $n$.

As the winding number $n$ increases further to $n \ge 5$, the positions of the minima of the field
$c = -1$ are shifted away from the symmetry axis forming a system of concentric rings. We observe that
the radii of these rings increase with increasing $n$, the number of rings and their structure
depending both on the winding number $n$ and on $m$, for example,  for a Skyrmion-anti-Skyrmion pair
a single ring in the $xy$ symmetry plane is formed \cite{Krusch:2004uf}.  For the configuration with $n=6,m=3$ we observe
one ring $(\rho_0^{(1)}=2.02, z_0^{(1)}=0)$ on the symmetry plane, and two other rings  placed symmetrically above and
below of it, $(\rho_0^{(2,3)}= 1.83, z_0^{(2,3)}= \pm 1.41)$.

Inclusion of the  pion mass term in the Lagrangian  \re{model} makes it possible for
$n=1$ Skyrmion-anti-Skyrmion
chains to exist. In Fig. \ref{f-3} we present the distribution of the topological charge density
and the isosurface of
the energy density for such a chain with $m=4$ at $\mu = 0.1$. Note that such solutions do not exist either
in Yang-Mills theory, nor in Skyrme theory in the absence of a pion mass potential. In these cases only
topological charge pairs of $n\ge 2$ and $n\le -2$ are found.

\begin{figure}
\lbfig{f-3}
\begin{center}
\includegraphics[height=.40\textheight, angle =-90]{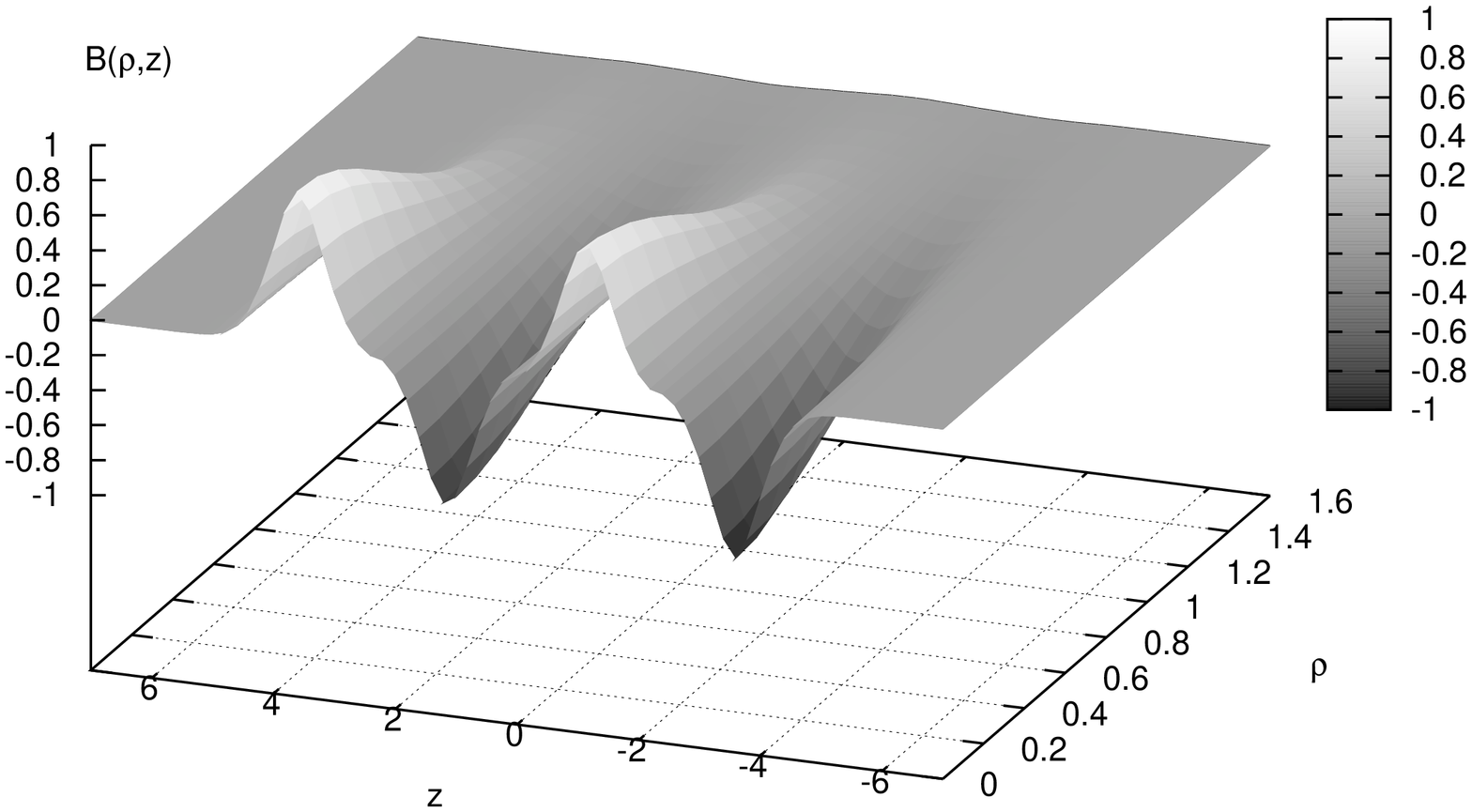}
\includegraphics[height=.45\textheight, angle =-90]{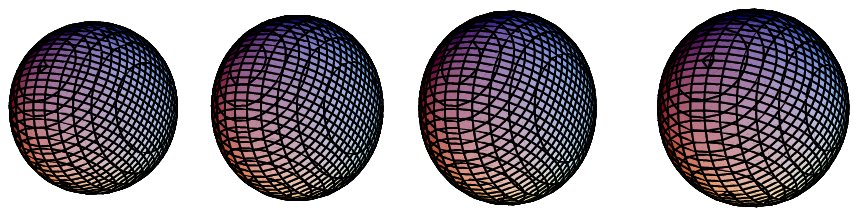}
\end{center}
\caption{
The topological charge density (top) $B(\rho,z)$ and the energy density isosurface
of the $m=4$ charge 1 Skyrmion and Skyrmion-anti-Skyrmion chain
are shown for $\mu = 0.1 $.
}
\end{figure}

An interesting analogy with the multi-monopoles (MM) and the monopole-antimonopole (MA)
chain solutions is that, the component of the Skyrme field
$\phi^4$ shows a clear relation to the corresponding behaviour of the magnitude $|\Phi|$ of the Higgs field of
MM's and MAP pair, respectively \cite{KKS}, as seen
in the plots in Fig ~\ref{f-2}. Indeed, these systems both are parametrised by two integer numbers, one of those
being associated with the topological charge of the constituents and
corresponds to the $n$-fold rotation about the symmetry axis, and the other winding number, $m$, appears
in the boundary conditions imposed on the field configuration. More specifically, in order to
construct the MA chains we impose asymptotic boundary conditions on the YMH system
which correspond to the $m$-fold rotation of the fields
as the polar angle $\theta$ varies from $0$ to $\pi$ \cite{KKS}.
Similarly, we impose the boundary condition $g(x,\pi) = m \pi$
on the angular function in the Ansatz (\ref{ansatz}) to construct the Skyrmion-anti-Skyrmion chains.
Furthermore, both in the YMH model and in the Skyrme model the structure of
the configurations changes as the winding number
$n$ increases beyond some critical value at which the positions of the constituents are no longer associated with
some set of isolated points on the symmetry axis but form circles around it.  Note that the similarity between
the monopole-antimonopole chains and chain solutions in the Skyrme model becomes even more transparent if we consider
axially symmetric YM caloron solutions where monopoles and antimonopoles are constituents of the caloron
\cite{Shnir:2007zz}. Then there is a holonomy operator
$\Tr {\cal P}(\vec r) = \cos || A_0(\vec r)||$ whose loci ${\cal P}(\vec r_0)=-1$ are associated with  the maxima of
the action density \cite{Shnir:2008ng}
.

\section{Conclusions}

We have constructed new static axially symmetric Skyrmion--anti-Skyrmion chain solutions numerically.
This was done by solving the two dimensional nonlinear partial differential equations as
a boundary value problem, using the same formalism and numerical techniques as were used in
the construction of the monopole--antimonopole chains~\cite{mapKK,KKS} and
non--self-dual instantons~\cite{Radu:2006gg}. In implementing the requisite boundary
conditions for this task, two integers $(m,n)$ are employed. Both these examples are
solved as $2$ dimensional problems, the former arising from the imposition of axial symmetry
in $\R^3$ and the latter of bi-azimuthal symmetry in $\R^4$. The integer $n$ labels the
topological charge of each constituent lump or anti-lump. In the monopole
case~\cite{mapKK,KKS} $n$ is the winding of the azimuthal angle in $\R^3$, while in
the instanton case~\cite{Radu:2006gg} it is the winding of the two azimuthal angles in $\R^4$,
taken to be equal. In both cases, it is the topological charge descendening from the second
Chern-Pontryagin charge subject to the respective symmetries. The integer $m$ enters the
asymptotic value of the (form factor) function which maps on to the remaining angular
coordinate, namely the first (polar) angle different from the azimuthal angle(s). $m$ is not
a topological charge. The total topological charge of any such finite action/energy
configuration vanishes when $m$ is {\it even}, and equals $n$ when $m$ is {\it odd}.

In the present work, we have restricted to $m=2, \dots , 6$.
Clearly, for $m=1$ we simply recover the axially symmetric MS of
charge $n$, which we have done as a warmup. We have examined the cases with values of $n$,
starting from $n=1$, through to $n=6$.

Our numerical investigations indicate fairly clearly that in the usual Skyrme model with no pion-mass term,
there exist no zero baryon number
solutions when each of the constituents carries baryon number $|n|<2$. However, inclusion of the pion-mass
term results in the existance of chains with $n=1$. This is perhaps not surprising in the background of the
known close analogy between solutions to Yang-Mills and sigma models respectively, $e.g.$, the usual YM model
and the Skyrme model with no pion mass term. This analogy is demonstrated most simply by considering the
unit charge solutions (BPST instantons) of the YM system parametrised by the radial function $w(r)$
on the one hand, and the unit charge solutions (hedgehog) of the Skyrme system parametrised by the radial
chiral function $f(r)$ on the other. This corresponce is $w=\cos f$, and it manifests itself already in the
one dimensional reduced actions of the two systems, but only in the absence of the pion mass potential
in the (Skyrme) sigma model. This is clear since the one dimensional reduced action does not feature
a pion mass like term $(1-\cos f)\equiv (1-w)$, in the presence of which the analogy between YM instantons and
Skyrmions disappears.

The $n=2, \mu=0 $ example was the one examined most intensively, since it is the first non marginal
case where we could reliably verify the existence of the Skyrmion-antiSkyrmion chains. However,
unlike in the two previous
examples in \cite{mapKK,KKS} and in \cite{Radu:2006gg}, the binding energy of the constituents
of the Skyrmion chain turned out to be quite weak, so the $n=2$ chains, especially for the odd numbers
of the consitutuents, are very unstable w.r.t. perturbations.

The nonexistence of the Skyrmion-antiSkyrmion chains with constituents carrying baryon numbers $\pm n$, for values
of $n\le n_{\rm min}$ is not surprising. In the corresponding analytic proof
\cite{Sadun:1992vj,Bor:1992ai} of existence for
non--self-dual instantons of zero Pontyagin charge, the existence of the case where the
constituents carried Pontyagin charge $n=|1|$ was not established, indicating its nonexistence.
In that case $|n_{\rm min}|=2$ which coincides with previous result of \cite{Krusch:2004uf}.

\bigskip
\noindent
{\bf\large Acknowledgements} \\

We would like to acknowledge numerous valuable discussions with Eugen Radu and Paul Sutcliffe.
This work was carried out in the framework of Science Foundation Ireland
(SFI) Research Frontiers Programme (RFP) project RFP07/FPHY330. Ya.S. is very grateful to the Department of
Mathematical Sciences, University of Durham for the hospitality in Durham.
\begin{small}

\end{small}

\end{document}